\documentclass[11pt,aps,nofootinbib,preprintnumbers,citeautoscript]{revtex4}

\usepackage{amsmath,amssymb}
\usepackage[dvipsnames]{xcolor}

\usepackage[colorlinks=true,citecolor=Red,linkcolor=Green]{hyperref}

\usepackage{enumerate}
\usepackage{graphicx}

\graphicspath{{/home/kapil/My_data/kapilpaper/arXiv/Robustness/}}
\usepackage{graphicx}
\usepackage{float}
\date{\today}
\begin{document}
\title{Quantum Adiabatic Feature Selection}

\author{Kapil K. Sharma$^\ast$ \\
\textit{Laboratory of Information Technologies,
Joint Institute for Nuclear Research,\\
6 Joliot-Curie St, Dubna, 141980, Russian Federation} \\
E-mail: $^\ast$iitbkapil@gmail.com
}

\begin{abstract}
Dimensionality reduction is the fundamental problem for machine learning and pattern recognition. During data preprocessing, the feature selection is often demanded to reduce the computational complexity. The problem of feature selection is categorized as a NP optimization problem. Exhaustive search of huge set of features takes huge amount of time on classical computer. In the present paper we discuss the role of quantum adiabatic computation to perform feature selection with bi-quadratic optimization and provide a quantum feature selection algorithm. Our algorithm runs with the quantum adiabatic time complexity bound $O(1/g_{min}^{2})$, which is better than classical approach for bi-quadratic feature selection.    
\end{abstract}
\maketitle

\section{Introduction}
Quantum machine learning\cite{qm1,qm2} is an active area of research which deals with handling the data on quantum computer and perform quantum algorithms. Development of varieties of algorithms in the context of quantum machine learning became exciting for last couple of years. In real world we deal with massive amount of data which carry the hidden information. To extract the useful information from data and its usage to train the machine is an art and involve huge diversities in statistical techniques. Performing any machine learning algorithm, the basic requirement is to reduce the dimensionality of the data and selection of the best features\cite{dm1}. Dimensionality reduction follow two approaches as feature extraction and feature selection\cite{dm2}. In feature extraction the set of existing features is transformed to lower dimensional space; principle component analysis belongs to this category\cite{d3}. While in feature selection the goal is to select the best features among the total subsets $2^{n}$ of the set of features $S$\cite{fs1}. The basic difference between feature extraction and feature selection is that; during the feature extraction the meaning of features changes while during feature selection the meaning of features remains same. In the present article we focus on feature selection problem, which is NP hard\cite{np1} and deal with combinatorial optimization. Feature selection has its deep importance in pattern recognition, image retrieval, gnomic analysis, machine learning, High energy physics (HEP) and others. However dimensionality reduction become much more important to reduce the number of features in real time pattern recognition to reduce the computation\cite{rfs1,rfs2}. For example the face recognition in real time with video stream; huge amount of pixels propagate to computer and feeding these in pattern recognition algorithm with preprocessing is infeasible. So, the real time feature selection has its own importance. Often dealing with feature selection problem, we are always interested to select the best features. Here the best features we mean the most distinguishable features; for example, given two classes of African and European people; it is obvious that European people have white skin while African people have black skin; here skin color is the most distinguish feature; of course there may be many another features which can also be taken into account. But most of the time the user apply the constraint on features to reduce the amount of computation. There are also other reasons for feature selection such as, a) to reduce the computational complexity if the data set is too large, b) to reduce the redundant features, c) to make the classification better. The redundant features are features which may act as noise or sometimes features are linearly dependent, on the other hand there is no use to take into account all the features which have linear relationship. Usually the easiest way to measure the relationship among features the Pearson  correlation coefficient (PCC)\cite{cc1} is used. If two features are independent than PCC is zero, but converse is not true. PCC being zero does not imply that features are independent, they may still be dependent. Hence PCC can not give the best information about the dependency of two random variables, it is long standing problem in statistics and often encountered in pattern recognition and machine learning as well. In the present work we use mutual information (MI)\cite{mi1}, which is always better than PCC. In classical domain, one approach for feature selection algorithms is to design an objective function and optimize by putting some constraints on feature subsets. The optimality of selected features have direct relationship to the error minimization of misclassification. However the definition of optimality can vary from problem to problem and one can set his own criteria of optimality while dealing with the machine learning problems. In other words the optimality can not be defined universally that can fit to every data set and it can always be criticized. As an example if there are $10^{2}$ features and one need to select the subsets of features with $10$ elements each, so $C^{10^{2}}_{10}=10^{12}$ optimal subsets are required, which is very huge number for searching. Even after searching it, there is no guarantee of optimality of the subsets obtained. Hence we require the fast algorithms to search whole space of $10^{12}$ subsets to select the optimal subsets. However it is sometimes better to avoid the exhaustive search as done in branch and bound algorithm\cite{bb1}. The algorithm is based on tree and use the property satisfied by the objective function, if the tree is too large the algorithms is difficult to implement. So, here we raise a question such that, can quantum computer tackle this problem of feature selection for Quantum machine learning algorithms\cite{em1}? In continuation of the discussion on feature selection approaches, we mention that two approaches are vastly adopted for feature selection such as forward feature selection and backward feature selection\cite{fb}. In forward feature selection approach; given an empty set of features, the goal is to keep adding the features at every step and test the decrement in the error, the process continues until further addition of feature significantly decreases the error. While in the backward feature selection approach; given a set of all features and remove the feature one by one and test the decrement in the error at every step, the process continues till the further removal of feature does not decrease the error significantly. There also exist a generalized algorithm in which one elements is added in the set and number of features are removed, it is called LR algorithm. Here we mention that, backward feature selection approach has good accuracy in comparison to forward feature selection approach as it takes into consideration the interdependency of features into account. We state that, these methods do not use the property of objective function and does not guarantee to provide the optimal feature subsets. In literature there are mainly three approaches adopted for feature selection such as, a) Filter method, b) Wrapper method, c) Embedded models\cite{dm1}. Each of these methods have its own performance criteria to judge the efficiency of the method. Filter method is easy to implement than other methods. The draw back of this method is that, it does not take into account the performance of the classifier. There are few good performance criteria used for filter method such as, a) Fisher score, b) Relief algorithm and c) Information gain\cite{fwe,wr1}. On the other hand wrapper method takes into account the performance of the classifier and have the greater degree of goodness over the filter method, but expensive in terms of computation. In the present article we deal with feature selection in supervised learning framework with quantum approach. Here we prefer filter approach with bi-quadratic optimization\cite{bo} performed with quantum adiabatic computation\cite{ac1}.  

\section{Framework of Quantum Adiabatic computation}
In this section, we discuss the model of adiabatic quantum computation (AQC) and its physics.  AQC is an analog quantum computation framework to solve combinatorial optimization problems\cite{ac2}, which also play the role to make the foundation of quantum annealing framework of computation\cite{ac3,ac4}. The basic idea of AQC is to prepare a physical system of spins which may follow the quantum Ising model. Initially the system is prepared with initial Hamiltonian $H_{0}$, which is non-diagonal in nature. Further the system is perturbed with a problem Hamiltonian $H_{p}$ which is the function of external magnetic field $B$ and interaction strengths among the spins $J_{i,j}$. Now the perturbed system evolve with respect to time, hence the eigenvalues and eigenvectors evolve with respect to time. It is expected for computation that the eigenspectrum of the system must not be same all the time. Hence to have different eigenvalues and eigenvectors of the system the following non-commutativity property must be satisfied,
\begin{equation}
[H_{0},H_{p}]\neq 0.
\end{equation}
The above condition imply that the Hamiltonian $(H_{0},H_{p})$ will not have same eigenvalues and eigenspecrtum. Suppose the perturbation is allowed with the time interval $0\leq t\leq T$, then one can write the convex combination of both the Hamiltonian w.r.t to time as,
\begin{equation}
H(s)=(1-s)H_{0}+sH_{p} \label{qa1}
\end{equation}
Where $s=\frac{t}{T}$ is the homotopy of $H_{0}$ to $H_{p}$. By using the unitary time evolution and following the Schrödinger equation, we can obtain the eigenstates and eigenvalues of the Hamiltonian $H(s)$. Given the increasing oder of eigenvalues as,
\begin{equation}
E_{0}\leq E_{1}\leq E_{2}\leq E_{3}\leq........\leq E_{n-1}
\end{equation}
We define, 
\begin{equation}
g(t)=E_{1}-E_{0} \label{dif1}
\end{equation}
To prevent energy level crossing in the system, the time evolution must be slow and adiabatic theorem\cite{at1,at2} must be satisfied. Defining the following,
\begin{eqnarray}
g_{min}=\min_{0\leq t\leq T} g(t); \quad g_{min}>0.
\end{eqnarray}      
Then the probability is obtained as,
\begin{equation}
||\langle \psi(0)|\psi (T)\rangle ||_{2}=1-\epsilon^{2} \label{c1}
\end{equation}
provided,
\begin{equation}
\frac{max||\frac{d}{dt}H(s) ||_{2}}{g_{min}^{2}}\leq \epsilon \label{c2}
\end{equation}
where $\epsilon \ll 1$. Both the above Eqs.\ref{c1},\ref{c2} imply that if $T\rightarrow \infty$ and $\epsilon\rightarrow 0$, then the delay factor can be defined as,
\begin{equation}
\gamma=\frac{max||\frac{d}{ds}H(s)||_{2}}{g_{min}^{2}}
\end{equation}
Typically the term $||\frac{d}{ds}H(s)||_{2}$ varies as a  polynomial with $n$. We are interesting to check the exponential changes in the time complexity, so the expression of the time complexity bound can be written as,
\begin{equation}
T=O\Big(\frac{1}{g_{min}^{2}}\Big)
\end{equation}
If the condition is satisfied as $T\gg \gamma$ then the quantum system will remain in the ground state and it defines the time complexity bound of global adiabatic quantum computation\cite{g1}. However Roland and Cerf\cite{g2}  suggested that the time complexity bound can be improved by using $\gamma$ as a instantaneous delay factor; means the factor $\gamma$ is the function of $s$. It is written as below,
\begin{equation}
\gamma(s)\approx \frac{||\frac{d}{ds}H(s)||_{2}}{g(s)^{2}}
\end{equation} 
with,
\begin{equation}
T\gg \int_{0}^{1}\gamma (s) ds 
\end{equation}
This is called local adiabatic quantum computation. The time complexity bound of this can be written as below,
\begin{equation}
T=O\Big(\int_{0}^{1}\frac{1}{g(s)^{2}} ds\Big)
\end{equation} 
The local adiabatic approach provides better time complexity bound but it is tough to find $g(s)$ and global adiabatic quantum approach is easy to implement because single parameter $g_{min}^{2}$ is required. In the present paper we adopt global adiabatic quantum computation.
\section{Formulating the Feature Selection Problem}
In this section we formulate the feature selection problem with bi-quadratic optimization by following the maximal relevance and minimum redundancy criteria (MRMR)\cite{mr1}. Here we use filter method, which does not depend on the performance of the target classifier. Let assume a set of features $\{x_{i}\}_{0\leq i\leq n}$ which form a feature vector $\{X=\sum w_{i}x_{i}\}$, the goal is to reduce the number of features and obtain a reduced set of features $\{x_{k<i}\}$ with MRMR. To measure MRMR we use an information measure criteria called as mutual information (MI). Here we recall that MI is a best measure in comparison to the Pearson correlation coefficient (PCC). Here we revise, if PCC is zero then it implies that two random variables are independent, but still these may be dependent. Hence PCC does not give the best information about the dependency of two random variables. If PCC is zero then still MI exists and prove that two random variables are still dependent. Hence in that sense MI capture the nonlinearity in the data and exists as a good measure of information. The mathematical expression of MI for discrete random variables is given as,
\begin{equation}
I(X;Y)=\sum_{x\in X}\sum_{y\in Y}P(x,y)\log \frac{P(x,y)}{P(x).P(y)}
\end{equation} 
If two random variables $(X,Y)$ are independent than $P(x,y)=P(x).P(y)$ and hence the factor $log\frac{P(x,y)}{p(x).p(y)}$ becomes $0$. Which shows surely two random variables $(X,Y)$ are independent. So to use MI we need to assign the weights $w_{i}$ to each feature $\{x_{i}\}$ with the feature vector $\{X\}$. These weights are treated as random variables. The corresponding probability of each weigh can be calculated as $(\frac{p}{q})$, where $p$ is the number of occurrence of a particular weight out of $q$. These probabilities are helpful to calculate the MI and to judge the MRMR. The classical bi-quadratic optimization function\cite{bo} for feature selection can be formulated as follows,
\begin{eqnarray}
Obj: \max_{X}X.(M_{n\times n}).X^{T}+\alpha(X-k)^{2} \label{op1} \\
\text{Subjected to}: m_{i,j}\geq 0\\
\sum_{i,j}^{n}m_{i,j}=1
\end{eqnarray}
where $M_{n\times n}$ is the MI matrix whose elements are $m_{i,j}$. The parameter $\alpha$ is the penalty strength to select the $k$ feature out of n features from the set $\{x_{i}\}_{0\leq i \leq n}$. Here we mention that the matrix $M_{n\times n}$ is a symmetric matrix, hence to make the calculations simple the diagonal part and upper triangular part can be taken into consideration as follows,
\begin{eqnarray}
M=M_{ii}+M_{i<j}; \quad 0\leq (i,j)\leq n\\
M_{ii}=I(X_{i};Y_{i}) \label{d1} \\
M_{i<j}=I(x_{i},x_{j}) \label{d2}
\end{eqnarray}
The diagonal elements of the matrix $M_{ii}$ equipped with the MI between the feature vector $\{X\}$ and the target classifier $\{Y\}$, while the upper triangular part of the matrix $M_{i<j}$ equipped with the MI among the probabilities corresponding to features $x_{i}$ and $x_{j}$.
\section{Quantum Adiabatic computation for feature selection} 
\begin{figure*}
\includegraphics[scale=1.15]{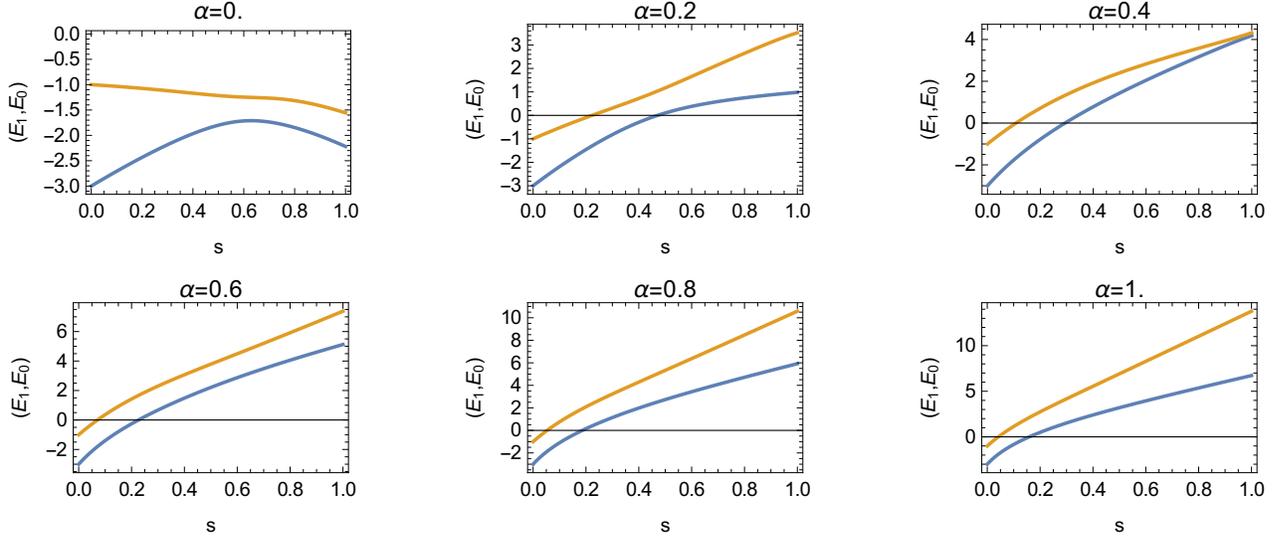} 
\caption{Plot of Eigenvalues w.r.t $\alpha$} \label{gmin}
\end{figure*}

\begin{figure*}
\includegraphics[scale=1.0]{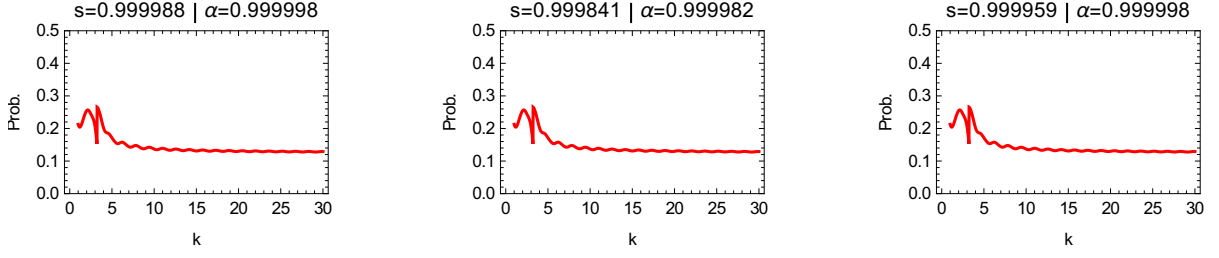} 
\caption{Probability of success $P(s^{\prime})$}\label{ps}
\end{figure*}

\begin{figure*}
\includegraphics[scale=1.0]{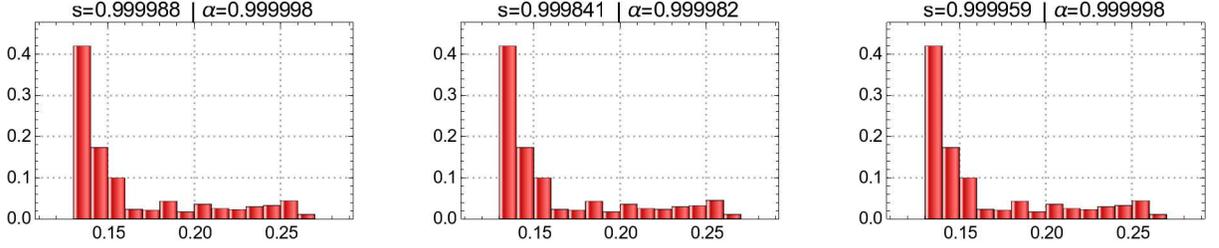}
\caption{probability distribution with $n=3$}\label{pdd}
\end{figure*}

In this section we encode the feature selection problem in quantum Ising model. A quantum Ising system is a collection of spins in space which have interactions among each other. The whole system is imposed with the external magnetic field. Generally multidimensional Ising model is very complicated so for simplicity 2D Ising model is very common in use, which is also considered in the present work. It is easy to formulate the problem in classical Ising system where the dynamics of spins are governed with the ``logical-and" operation among the spins ie. $(s_{i}.s_{j})$ . While on the other hand for quantum Ising model, each spin is dealing with the Pauli spin operator and Kronecker tensor product among all the spins ie. $(\sigma_{i}\otimes \sigma_{j})$. By following the stranded procedure we can encode the feature selection problem presented in Eq.\ref{op1}, in problem Hamiltonian $H_{p}$ by using quantum Ising model. The problem Hamiltonian $H_{p}$ can be fragmented in two parts as quantum Ising Hamiltonian and penalty Hamiltonian as given below.
\begin{eqnarray}
H_{p}=H_{qih}+H_{penalty}\\=\underbrace{-\displaystyle\sum_{i}^{n}b_{i}\sigma^{i}_{z}}_{M_{ii}}
-\underbrace{\displaystyle\sum_{i< j}^{n}J_{i,j}\sigma^{i}_{z}\sigma^{j}_{z}}_{M_{i<j}}
+\underbrace{\alpha \displaystyle\sum^{n}_{i}(\sigma^{i}_{z}-k.I_{n\times n})^{2}}_{penalty}\label{m1}
\end{eqnarray}
Here $b_{i}$ is the magnetic filed corresponding to each qubit which encodes the bias of each individual qubits. And $J_{i,j}$ is the interaction strength between the spins $i$ and $j$ with $(i\neq j)$. In Eq.\ref{m1}, the first and second terms encode the information available in Eqs.\ref{d1} an \ref{d2} respectively. The problem Hamiltonian $H_{p}$ is required to schedule the quantum adiabatic computation as per the Eq.\ref{qa1}.  Following the Eq.\ref{qa1}, we choose the initial Hamiltonian $H_{0}$ such that it must not commute with $H_{p}$, 
\begin{equation}
H_{0}=\displaystyle\sum_{i}^{n}\sigma_{x}^{i}; \quad [H_{0},H_{p}]\neq 0.
\end{equation}
To start the adiabatic quantum computation, we prepare the initial quantum state of Ising system as equal superposition of $n$ number of qubits as,
\begin{eqnarray}
|\psi(0)\rangle=\frac{1}{\sqrt{N}}\displaystyle\sum_{i=0}^{n}|i\rangle; \quad \quad i\in \{0,1\}^{n}.
\end{eqnarray} 
After applying the perturb Hamiltonian $H_{p}$ to the initial Hamiltonian $H_{0}$, the time evolution of the system can be obtained as per the Schrödinger time dependent equation. It is given as,
\begin{equation}
|\psi(s)\rangle=e^{-\frac{i}{\hslash}\int_{0}^{s^{\prime}} H(s) ds}|\psi(0)\rangle
\end{equation} 
Here we consider the homotopy scheduling time as a unit time interval $s^{\prime}\in[0,1]$ to make the simulation easy. In the present work our goal is to select the $k$ best features out of $n$ features. So in general we need $n$ number of qubits to encode $2^{n}$ features, but for better encoding strategy, it is always preferable to encode more number of features in less number of qubits to make the quantum computation cost effective. So in this direction we choose the condition $(n/k)<1$. As time evolve we can calculate the factor given in Eq.\ref{dif1} at time $s^{\prime}$. Measuring the ground state of $H(s^{\prime})$ gives the solution of the problem. Let suppose the structure of the ground state is obtained as $(0000......f_{i}.....000000)$, if $f_{i}$ is $1$ then $t^{th}$ feature is selected. If the structure is $(0000......f_{i}.0000..f_{j}......000000)$, if $(f_{i},f_{j})$ are $(1,1)$ then $(i,j)^{th}$ features are selected; and so on. Simultaneously we can calculate the probability of success at time $s^{\prime}$ as,
\begin{equation}
p(s^{\prime})=||\langle \psi(0)|\psi(s^{\prime}\rangle||^{2}
\end{equation}  
Here the probability of success $p$ is the function of the parameters $(n,k,\alpha)$. In our simulation work we sample the random variables $w_{i}$ following the Normal  Distribution with  zero mean and variance one and further calculate the MI. We fix the seed on the machine to generate these random variables so that repetition of the program produces the same random numbers on the same machine. However in realistic hardware of quantum annealer these values differ. D'Wave quantum annealer platform has its own sampler\cite{dw}. But for analytical study to observe the theoretical time complexity of the algorithm, it is must to generate the random variables $w_{i}$ on classical machine. Here we show some simulation results, in that direction we plot the eigenvalues $(E_{0},E_{1})$ w.r.t to parameter $\alpha$ for $n=3$ qubits and $k=5$ features in Fig.\ref{gmin}. Here we follow the condition $(n/k)<1$ for better encoding.
Further we gives the simulation results for $(g_{min})$ with $n=3$ qubits in the following table.\\
\begin{center}
\begin{tabular}{c c c c c}
\hline 
k & s & $\alpha$ & $g_{min}$ \\ 
\hline 
5 & 1 & 1 & -7.05986 \\ 
\hline 
10 & 1 & 1 & -27.0599 \\ 
\hline 
15 & 1 & 1 & -47.0599 \\ 
\hline 
20 & 0.999988 & 0.999998 & -67.0592 \\ 
\hline 
25 & 0.999841 & 0.999982 & -87.0477 \\ 
\hline
30 & 0.999959 & 0.99998 & -107.055 \\ 
\hline 
\end{tabular} 
\end{center}
In above table we observe as the number of features $(k)$ increases,  the minimum energy gap $(g_{min})$ is achieved with the upper bound of time complexity $s=O(0.999)$, as the number of features increases the time decreases. We also plot the probability of success $p(s^{\prime})$ in Fig.\ref{ps} with $(k=30)$ features. We have found that for less number of features the probability of success $p(s^{\prime})$ is high, as the number of features increases the probability decrease but almost remains constant for large number of features. Hence algorithm performs better for less number of features. We also show the probability distribution in the figure \ref{pdd}. Here we recall that, the upper bound of the time complexity of adiabatic quantum computation is given by $O(1/(g_{min})^{2})$. It is plotted for $n=\{3,5,7\}$ qubits in Fig.\ref{qcp1}. We have found as number of qubits increases, the growth rate of the function remain asymptotic and increasing number of qubits only effect the ability to select large number of features.  

\section{The algorithm and performance comparison}
\begin{figure*}
    \centering
    \begin{minipage}{0.45\textwidth} 
    \centering
       \includegraphics[scale=0.85]{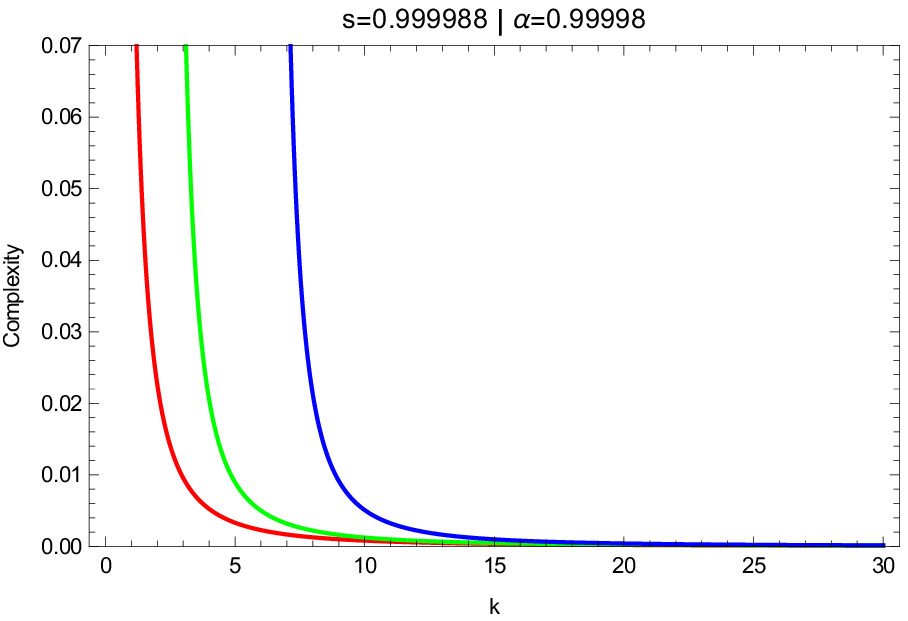}
        \caption{quantum complexity plot}\label{qcp1}
    \end{minipage}
    \begin{minipage}{0.45\textwidth} 
    \centering
        \includegraphics[scale=0.85]{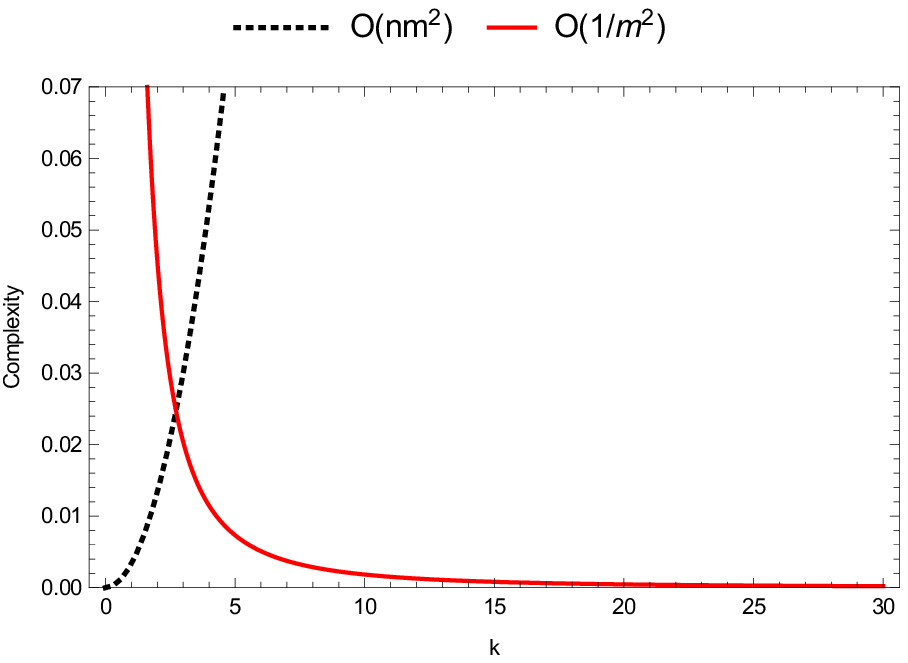}
        \caption{quantum complexity plot}\label{qcp2}
 \end{minipage}    
\end{figure*}

Here we present our quantum adiabatic features selection algorithm.
\begin{itemize}
\item Start: Empty Set $E=\{Null\}$.
\item Set of initial features: $I=\{n\}$.
\item Schedule AQC: $s=0$.
\item Find Ground state of $H(s^{\prime})$; Select $\{x_{i}\}$.
\item Fill $E$ as: $x_{i}\rightarrow S$; End $s=T$.
\end{itemize}
We have been calculated the upper bound of adiabatic time complexity as $O(0.999)$, which represent the fastest time to select the minimum number of features. Following the work in Ref.\cite{bo} the classical computational complexity has been calculated as $O(nm^{2})$, where $n$ is the training size of the data and $m$ is the number of features. We have been plotted the classical computational complexity and adiabatic time complexity bound in Fig.\ref{qcp2}. The black color graph shows the classical complexity and red color graph shows the quantum complexity. We have found that the growth rate of the function $O(1/m^{2})$ is better than $O(nm^{2})$ for less number of features and hence quantum complexity is always better for feature selection.  

\section{Conclusion}
In the article we have shown the role of adiabatic quantum computation for feature selection problem. The problem of feature selection is very important for pattern recognition and machine learning tasks. As the domain of quantum information and computation is quite active, so feature selection on quantum computer for quantum pattern recognition and machine learning tasks is obliviously interesting. In the present work show the efficacy for quantum approach to select the best features with the upper bound of time complexity having the order as $O(0.999)$, which better to select minimum number of features in comparison to classical one. The given order of complexity follow the asymptomatic nature for adiabatic quantum computation ie. $O(1/g_{min}^{2})$. Further we have shown as the number of qubits increases to encode the features, the growth rate of bound of time complexity remain same and still this bound is the order of $O(0.999)$, the increasing number of qubits only effect the encoding efficiency of features. The present work can be enhanced for further feature selection problems related to another framework such as wrapper and embedded methods. We are in hope the work may be useful for quantum machine learning community.

\end{document}